# Improving FLAIR SAR efficiency at 7T by adaptive tailoring of adiabatic pulse power using deep convolutional neural networks

Shahrokh Abbasi-Rad[1], Kieran O'Brien[2,3], Samuel Kelly[1], Viktor Vegh[1,3], Anders Rodell[2], Yasvir Tesiram[1], Jin Jin[2], Markus Barth[1,3*], Steffen Bollmann[1,3*]

[1]*Centre for Advanced Imaging, the University of Queensland, Brisbane, Queensland, Australia.*
[2]*Siemens Healthcare Pty Ltd, Brisbane, Queensland, Australia.*
[3]*ARC Training Centre for Innovation in Biomedical Imaging Technology*
[*]*Steffen Bollmann and Markus Barth share senior authorship role.*

**Correspondence to:**

Shahrokh Abbasi-Rad

The University of Queensland

Centre for Advanced Imaging

Building 57, Research Road, St. Lucia

Brisbane, QLD, 4072

Australia

Email: s.abbasirad@uq.edu.au

# Abstract

**Purpose**: The purpose of this study is to demonstrate a method for Specific Absorption Rate (SAR) reduction for T2-FLAIR MRI sequences at 7T by predicting the required adiabatic pulse power and scaling the amplitude in a slice-wise fashion.

**Methods**:

We used a TR-FOCI adiabatic pulse for spin inversion in a T2-FLAIR sequence to improve $B_1^+$ homogeneity and calculate the pulse power required for adiabaticity slice-by-slice to minimize the SAR. Drawing on the implicit $B_1^+$ inhomogeneity present in a standard localizer scan, 3D AutoAlign localizers and SA2RAGE $B_1^+$ maps were acquired in eight volunteers. A convolutional neural network (CNN) was then trained to predict the $B_1^+$ profile from the localizers and scale factors for the pulse power for each slice were calculated. The ability to predict the $B_1^+$ profile as well as how the derived pulse scale factors affected the FLAIR inversion efficiency were assessed in transverse, sagittal, and coronal orientations.

**Results**:

The predicted $B_1^+$ maps matched the measured $B_1^+$ maps with a mean difference of 4.45% across all slices. The acquisition in the transverse orientation was shown to be most effective for this method and delivered a 40% reduction in SAR along with 1min and 30-sec reduction in scan time (28%) without degradation of image quality.

**Conclusion**:

We propose a SAR reduction technique based on the prediction of $B_1^+$ profiles from standard localizer scans using a CNN and show that scaling the inversion pulse power slice-by-slice for FLAIR sequences at 7T reduces SAR and scan time without compromising image quality.

# 1. Introduction

T$_2$-weighted MR imaging of the brain visualizes subtle brain lesions and is a crucial contrast for clinical applications. The commonly used T$_2$-weighted Fluid-Attenuated Inversion Recovery sequence (T$_2$-FLAIR) (1) further increases both the conspicuity and detection of lesions and avoids cerebrospinal fluid (CSF) artifacts from partial volume averaging and fluid motions during the cardiac and respiratory cycles (2, 3). The T$_2$-FLAIR sequence is utilized for imaging in a wide spectrum of diseases such as infections, white matter diseases, tumors, vascular diseases, and multiple sclerosis (2, 4-6).

7T scanners offer a high signal-to-noise ratio allowing the improved visualization of anatomical details in white matter, such as the optic radiation and subnuclear structures in the thalamus (7-10). However, at ultra-high-field, the wavelength of the radiofrequency field is comparable to the dimensions of the object (11) causing non-uniform spin excitation, refocusing or inversion that can degrade image quality. Therefore, implementing FLAIR at 7T creates additional challenges due to transmit field $B_1^+$ inhomogeneities caused by both local field-tissue interactions and large-scale effects such as the elliptical eccentricity of the head (12-14). Non-uniformities in transmit $B_1^+$ field result in spatially variant inversion efficiency and lead to incomplete CSF suppression. Regions mainly affected by this include the base of the skull and the temporal lobes of the brain.

The adiabatic full passage (AFP) generates inversion of the spins depending on the variation of amplitude and frequency modulation functions of B$_1$ (15). Once the adiabatic condition is satisfied, the inversion efficiency will be uniform throughout the whole slice and the flip angle is independent of the pulse power (16). Thus, adiabatic pulses are robust to the inhomogeneous profile of the transmission field (17) and are commonly "over-driven" to always ensure inversion at the cost of increasing the specific absorption rate (SAR) of the sequence. To satisfy SAR limitations at ultra-high field, compromises are made, such as acquiring fewer slices, prolonging the repetition time (TR) or adding a delay time after each scan, leading to reduced coverage and/or increased scan times.

This shows that it is challenging to improve $B_1^+$ homogeneity, stay within the SAR limits and achieve a high image quality at 7T. Recently, several attempts for optimizing adiabatic inversion have been made: For 3D acquisition protocols with non-selective inversion,

O'Brien *et al.* used the combination of adiabatic RF pulses and high permittivity dielectric pads to reach high-quality images with high inversion efficiency and acceptable SAR limits, because the high permittivity of the pads allowed lowering the power of the pulse (18). However, the optimal placement of the pads around the subjects' head is difficult to achieve. Beqiri *et al.* used an 8-channel transmit head coil and direct signal control optimization for dynamic RF shimming on a pulse-by-pulse basis through the echo train length (19). Gras *et al.* designed $k_T$-point pulses based on previously acquired $B_1^+$ and $\Delta B_0$ maps from subjects and showed comparable homogeneity to the adiabatic version but with significantly lower SAR (20). Nonetheless, the heterogeneous SAR profiles resulting from each channel are difficult to model and the techniques are limited to non-selective 3D inversion.

For multi-slice 2D protocols, such as a standard $T_2$-FLAIR acquisition, the slice selective adiabatic inversion pulse is a major contributing factor to the overall SAR of the sequence and limits the flexibility afforded to the other RF pulses (excitation and refocusing) used in the sequence. We propose that a substantial SAR reduction could be achieved by adapting the pulse power on a slice-by-slice basis from $B_1^+$ profiles in image slices and reduce the need for "over-driving" the RF power. This would ensure that in each slice only the RF power necessary for the adiabatic condition is being used. However, this means that one needs information on the spatial $B_1^+$ distribution within the whole volume, and this is commonly done by acquiring a $B_1^+$ map, something that results in additional scan time and influences the workflow, counteracting the possible scan time reduction gained from reducing SAR. We have shown previously (21) that localizer scans that need to be acquired anyways, are implicitly sensitive to $B_1^+$ inhomogeneities and that it is possible to extract sufficiently accurate $B_1^+$ information from these scans to predict the RF power for adiabatic pulses needed for 2D $T_2$-FLAIR imaging at 7T. Altogether, the main aim of this work is to reduce SAR requirements of $T_2$-FLAIR imaging at ultra-high field MRI by tailoring the power of the inversion pulse in a slice-by-slice fashion while maintaining image quality for the three commonly used slice orientations (transverse, sagittal, and coronal). In order to obtain the necessary $B_1^+$ information without prolonging measurement we estimated the scale factors for RF pulse power scaling by predicting the $B_1^+$ map using a convolutional neural network (CNN) using the localizer as input, which we verified by direct measurement of the $B_1^+$ map using a SA2RAGE (22). Importantly, to achieve robust scale factor estimation we investigated the adiabaticity of the AFP pulse for each slice by determining those areas where CSF suppression might not be achieved due to inadequateness of the $B_1^+$ field and compared the image quality with the images acquired using full RF power.

## 2. Methods

### 2.1. MR Imaging

Imaging was performed on a 7T whole-body research scanner (Siemens Healthcare, Erlangen, Germany), with maximum gradient strength of 70 mT/m and a slew rate of 200 mT/m/s. A 7T 1-channel Tx 32-channel Rx head array coil (Nova Medical, Wilmington, MA, USA) was used for radiofrequency transmission and signal reception. Third order shimming was employed to improve the $B_0$-field homogeneity.

For $T_2$-FLAIR imaging the time resampled frequency offset independent (TR-FOCI) pulse, which was shown to outperform HS8 and FOCI adiabatic RF pulses (18, 23, 24), was used to invert the spins despite the strong $B_1^+$ inhomogeneities. The sequence was applied in two different modes: a) in "non-scaled mode" the power of the adiabatic pulse was the same for all slices; and, b) in the "slice-by-slice scaled mode" where the power of the inversion pulse was scaled for each slice according to the scale factors calculated from either the measured or the predicted $B_1^+$ profiles. Except for the pulse power, all other parameters were kept the same.

We conducted three main imaging experiments: i) $B_1^+$ Map and 3D localizer data for training the CNN ii) validating the $B_1^+$ profile prediction of the CNN; and iii) the main SAR reduction and image quality validation of $T_2$-FLAIR.

The imaging protocol and parameters for the experiments were:

1. <u>AutoAlign 3D localizer</u> acquired using a gradient-echo sequence with the following parameters: TA=15.74s, TR=4ms, TE=1.53ms, α=16°, matrix=160x160x128, FOV=260x260x260mm³, GRAPPA=3.
2. Individual $B_1^+$ profiles, acquired using the <u>SA2RAGE</u> sequence (22) with the following parameters: TA=1min53s, TR=2.4s, TE=0.93ms, α=6°, $TI_1$=108ms, $TI_2$=1800ms, matrix=64x64x64, FOV=288x288x288mm³. 6/8 partial Fourier encoding in the phase direction was used to reduce $TI_1$ for increasing the $T_1$ insensitivity of the method.

3. To obtain slice position information for the FLAIR scan, a fast 3D Gradient Echo scan (9 seconds) was acquired before the FLAIR sequence with identical geometry. The sole purpose of these images was to get the positional information of the slices for the reslicing step. <u>Gradient echo</u> images were acquired three times with three orientations with the following parameters: TA=9.4s, TR=218ms, TE=2.28ms, α=5°, matrix=64x54x40, FOV=224x224x144 mm$^3$, GRAPPA = 2.
4. <u>Standard whole-brain FLAIR</u> images (all three orientations) were acquired with the following parameters: TA=3min18s, TR=9s, TE=100ms, TI=2.6s, α=150°, ETL=9, slices=40, thickness=3mm, matrix=320x256, FOV=223x179mm$^2$, GRAPPA=3.

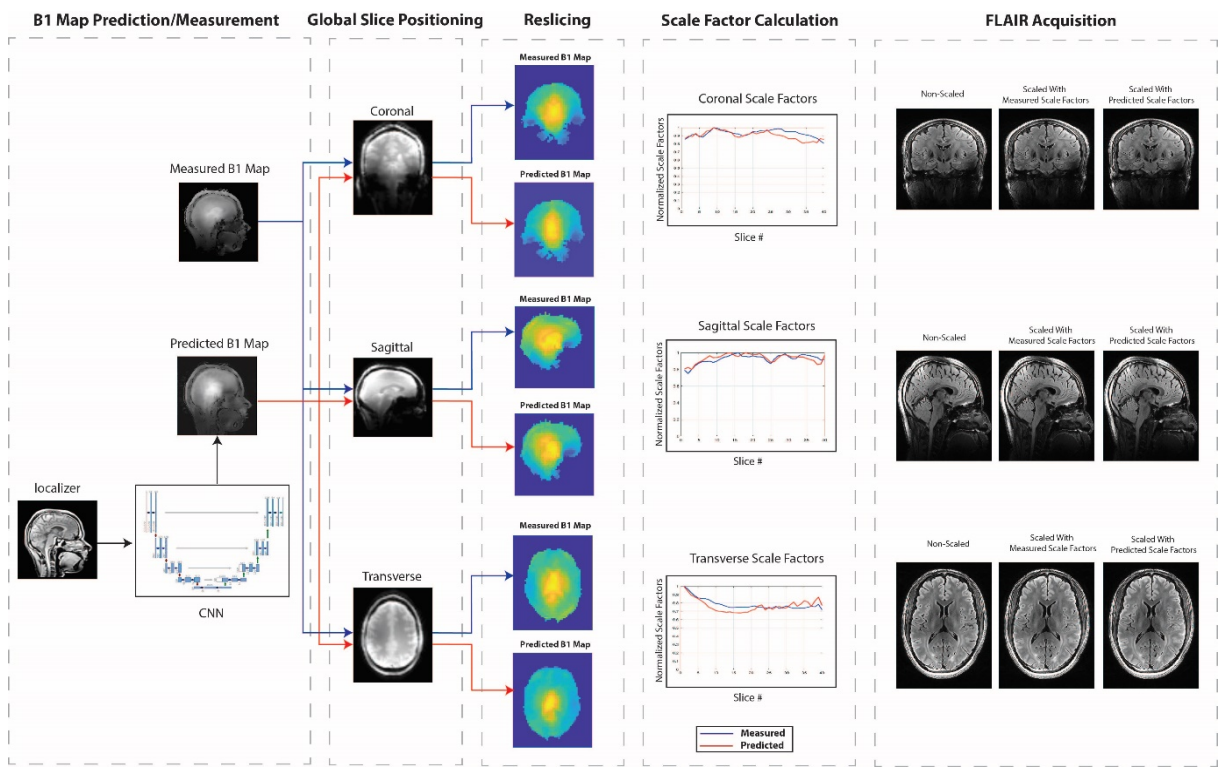

Figure 1. A flow diagram of the whole process of scale factor calculations based on both actual and predicted $B_1^+$ profiles.

**Figure 1** shows a block diagram of the experiment: first, five images were acquired (localizer, SA2RAGE $B_1^+$ profile, three fast GRE images for slice positioning); second, processing was performed while the subjects were still in the scanner to calculate the scale factors; third, nine T$_2$-FLAIR images were acquired as follows: Three non-scaled (standard) FLAIR acquisitions, three FLAIR images scaled slice-by-slice by scale factors calculated from measured $B_1^+$ profiles and three FLAIR images scaled slice-by-slice by scale factors calculated from predicted $B_1^+$ profiles. The different parts of the method are described in detail as follows.

## 2.2. Participants

We obtained written informed consent from participants prior to *in vivo* scanning as approved by the local human ethics committee. All procedures performed in studies involving human participants were in accordance with the ethical standards of the institutional and/or national research committee and with the 1964 Helsinki declaration and its later amendments or comparable ethical standards. Twenty-six volunteers participated in this study and a subgroup of these have been scanned for each of the three experiments.

Image data of a group of ten participants (6M/4F) was used for training, validation, and evaluation of the robustness of the CNN predictions (using AutoAlign (scan 1) and SA2RAGE (scan 2)). Pairs of images of localizer and SA2RAGE $B_1^+$ profiles were acquired. Data of eight participants were used for CNN training (seven) and testing (one). Thereafter, data of the remaining two participants was used for validating the CNN prediction performance for different head rotation scenarios (Section 2.4.2).

Image data of the second group (ten participants (4F/6M)) were used to calculate the scale factors in three primary imaging orientations (transverse, sagittal, and coronal) using the two different strategies, measured (using SA2RAGE (scan 2)) and predicted (using AutoAlign (scan 1) and the CNN), and investigate the agreement of adiabatic pulse power prediction between them.

Image data of the third group (six participants (3M/3F)) were used to test the SAR reduction and image quality. FLAIR images (using FLAIR (scan 4)) in both standard (non-scaled) and slice-by-slice scaled modes, in three different orientation were acquired.

## 2.3. Convolutional Neural Network

### 2.3.1. CNN Architecture and Training

The Convolutional Neural Network (CNN) was implemented in Python 3.6 using tensorflow and tensorboard v1.13 (25) and Keras v2.2.4 (26) and was trained on an NVIDIA Tesla K40c graphics card. The network is based on a modified version of the established 3D U-Net architecture (27). The contracting part consists of three-dimensional convolution layers with filters of size 3x3x3, a stride length of 1x1x1, rectified linear units (ReLU) and pooling layers.

The expanding part of the network consists of transposed convolutional layers with filters of size 2x2x2, a stride length of 2x2x2 and ReLUs, and convolutional layers identical to the contracting part. The architecture also utilizes skip connections.

For network training, we acquired AutoAlign 3D localizer and B1map with the matrix sizes of 160 x 160 x 128 and 64 x 64 x 64, respectively from 8 subjects. We used seven datasets for training and one dataset for testing. We resliced (up-sampled) the $B_1^+$ map into the localizer imaging volume and randomly cropped 1000 samples with the size of 32x32x32 from each image. The network was trained on 6000 patches and tested with 1000 patches selected randomly from the localizer and $B_1^+$ data, which were masked using Brain Extraction Tool (BET) (28) to exclude non-brain tissues. The CNN was trained for 200 epochs with a learning rate of 0.02, a batch size of 16 and 20 percent validation data.

### 2.3.2. CNN head orientation robustness evaluation

To test the performance of the network in situations with head positions different from the training data, we examined two subjects in five different head positions. The volunteers rotated their head into four extreme scan-positions from baseline: left, right, front, and back. For each position of the head, we acquired a localizer and predicted the $B_1^+$ profile using the trained network. We then used the acquired SA2RAGE $B_1^+$ map to calculate the relative error of the network prediction.

The error was calculated pixel-wise according to the equation below and an error map was generated.

$$prediction\ error = \frac{predicted\ B1 - measured\ B1}{Measured\ B1} \times 100 \quad \text{eq. 1}$$

### 2.4. Reslicing

The purpose of the SAR reduction experiment for FLAIR acquisitions was to scale the adiabatic pulse power by calculating slice-specific scale factors. Prior to the scale factor calculation, acquired FLAIR slice positions have to be determined. As a proof of concept, we achieved this by acquiring a fast (9 sec) gradient-echo sequence with the exact slice positions required for the FLAIR sequence. The predicted and measured $B_1^+$ profiles were then resliced into the desired imaging volume using the SPM12 package in MATLAB 2018b (Mathworks).

## 2.5. Absolute B1 map and Adiabatic Threshold

To perform a pixel-wise investigation of whether the spins would be inverted given the achieved $B_1^+$, we calculated the absolute $B_1^+$ map for the inversion pulse:

$$B_1^{Inv} = B_1^{ref} \frac{V^{op}}{V^{ref}} \quad \text{eq. 2}$$

where the SA2RAGE $B_1^+$ map was considered the reference scan (i.e., ref). Using the fixed reference voltage amplitude for both scans, the ratio of the operational amplitude of the inversion (i.e., inv) pulse ($V^{op}$) to the voltage amplitude used for SA2RAGE ($V^{ref}$) yields the factor for converting the relative $B_1^+$ map to the absolute $B_1^+$ map.

To calculate the absolute $B_1^+$ value required for the adiabatic condition and to determine the adiabaticity threshold, the full Bloch equations were simulated for an inversion TR-FOCI pulse with a slice thickness of 3mm and pulse length of 20 ms as used in the FLAIR sequence protocol.

The amplitude, frequency, and gradient modulation functions for the applied TR-FOCI pulse are shown in Figure 2a and the slice profile of the pulse simulated by numerically solving the Bloch equations is shown for a range of maximum amplitude values for an amplitude modulation function from zero to 200 Hz (Figure 2c). These simulations suggest that for an inversion efficiency of more than 97%, a 150 Hz amplitude transmit field is required (Figure 2b.) Therefore, CSF suppression is achieved where the $B_1^+$ value was higher than the simulated nominal value.

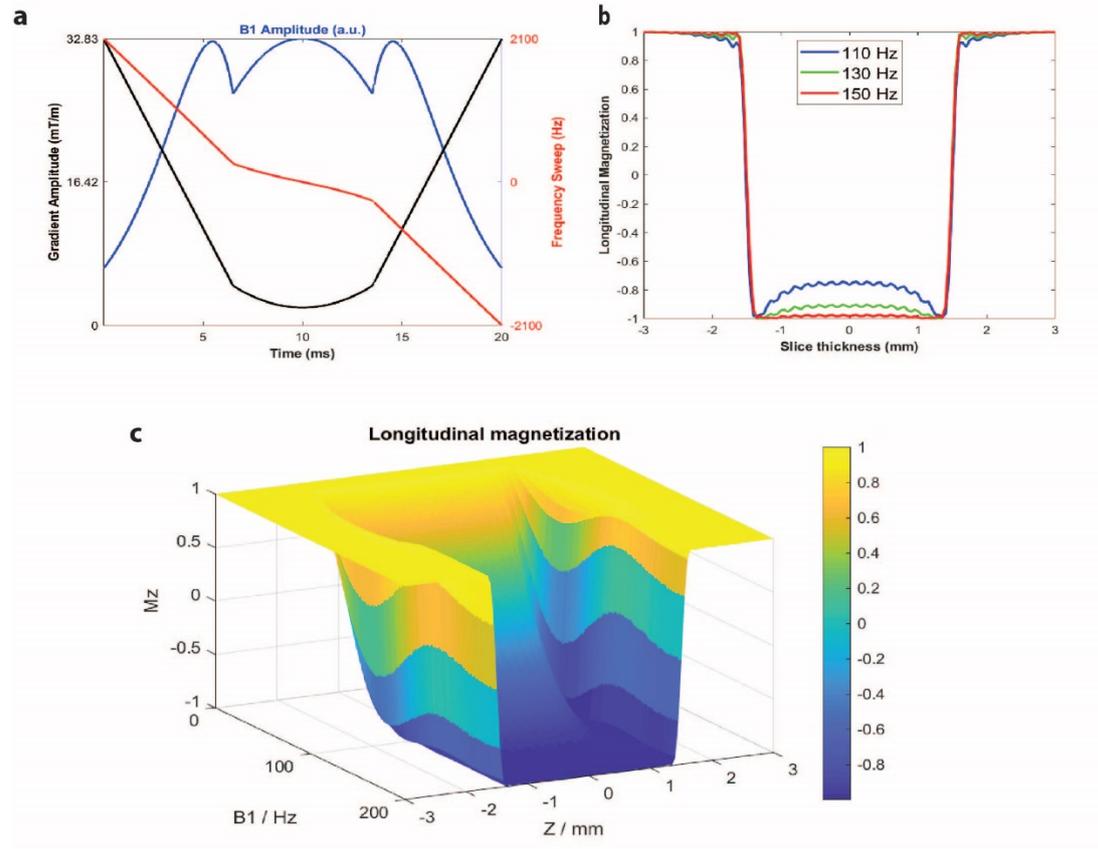

*Figure 2.* The results of the full Bloch equation simulation for the inversion pulse used in the T2-FLAIR sequence in this study. Modulation functions for amplitude (blue), frequency (red), gradient (black) are shown (a) for a TR-FOCI pulse with the length of 20 ms inverting slices with the thickness of 3 mm. Shown also are b) the slice profiles with different $B_1^+$ amplitudes, and c) that adiabaticity is satisfied from the amplitude of 150 Hz and held for the amplitudes above the threshold.

## 2.6. Scale factors

For scale factor calculation, we first removed non-brain voxels using the brain extraction tool in FSL (28). All voxels above the adiabatic $B_1^+$ threshold calculated in the previous section were then excluded. We then selected the upper bound of the 95% confidence interval around the mean for each slice and used the equation below for calculating the slice scale factor:

$$\text{Scale Factor} = \frac{Nominal\ B1\ Value}{Upper\ Bound\ B1\ Value} \quad \text{eq. 3}$$

with the "*Nominal B1 value*" the $B_1^+$ value for the adiabatic condition to be satisfied, which is calculated from the reference transmitter voltage used for RF calibration and the TR-FOCI pulse parameters including pulse length and slice thickness.

## 2.7. SAR reduction Index

To calculate the SAR reduction index for the FLAIR acquisition in scaled-mode we use equation 4 as below:

$$\text{SAR} = \frac{\sigma|E^2|}{2\rho} \quad \text{eq. 4}$$

where $\rho$ the density and $\sigma$ is the tissue's electrical conductivity, and E is the electrical field. To calculate the accumulative SAR for the whole experiment we added the SAR value for all slices and used the scale factors for the adiabatic pulse power for each slice. Therefore, we can rewrite equation 4 as:

$$SAR_{Scaled} = \sum_{i=1}^{N_{slc}} \frac{\sigma|(k_i.E)^2|}{2\rho} \quad \text{eq. 5}$$

where, $N_{slc}$ is the number of slices and $k_i$ is the scale factor for the i*th* slice. To measure the SAR reduction factor we did the experiment twice: For the standard mode, we used the scale factor of one for all slices and for the scaled-mode we used the calculated scale factors. By dividing the SAR computed for the scaled mode by the one computed for the non-scaled mode the SAR reduction index was computed as:

$$\text{SAR reduction index} = \frac{SAR_{Scaled}}{SAR_{Non-Scaled}} = \frac{\sum_{i=1}^{N_{slc}} \frac{\sigma|(k_i.E)^2|}{2\rho}}{\sum_{i=1}^{N_{slc}} \frac{\sigma|E^2|}{2\rho}} = \frac{\frac{\sigma|E^2|}{2\rho}\sum_{i=1}^{N_{slc}} k_i^2}{\frac{\sigma|E^2|}{2\rho} N_{slc}} = \frac{\sum_{i=1}^{N_{slc}} k_i^2}{N_{slc}} \quad \text{eq. 6}$$

This SAR reduction index shows the reduction of the SAR only related to the adiabatic inversion pulse, which differs from the scanners' SAR look ahead monitor as this value is for the whole sequence, which in our case has nine other refocusing RF pulses.

# 3. Results

Figure 3 shows five different slices of the localizer as the input to the trained CNN, the measured $B_1^+$ map as the ground truth, the predicted B1 map as the output of the network, as well as the relative error map calculated using equation 1. The predicted $B_1^+$ map mostly resembles the measured $B_1^+$ map. Although there are several voxels in the error map where the network over- or under-predicted the $B_1^+$ values (maximum up to 60%), the mean difference between predicted and measured $B_1^+$ map across all slices is 4.46%.

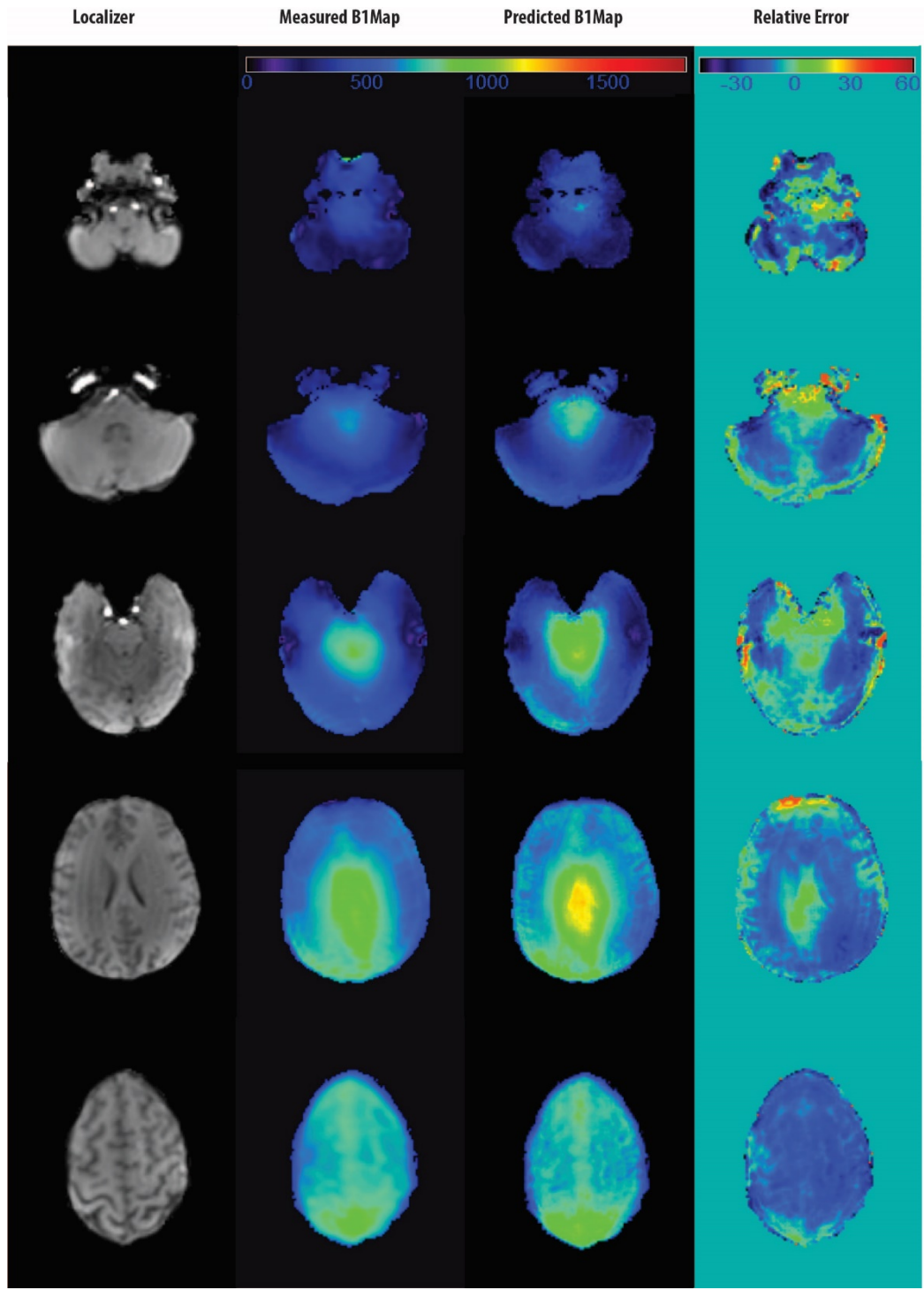

***Figure 3.*** Shown are the predicted $B_1^+$ maps versus the measured maps. From left to right: localizer, measured $B_1^+$ map acquired with SA2RAGE sequence, predicted B1 map, and the relative error of prediction. The measured $B_1^+$ map was up-sampled (resliced) to the localizer scan. From top to bottom, the slices show inferior to superior sections across the brain. The data is from a 27-year-old male participant.

The results of the CNN prediction for one example subject in five different head positions are shown in Figure 4. The localizer scans, first (axial view) and second (sagittal view) columns, show the axis of the head rotation in each scenario. Predicted $B_1^+$ maps, measured $B_1^+$ maps, as well as error maps are shown for the mid-slice in the third, the fourth and the fifth columns, respectively. The predicted maps in different head positions indicate that the prediction error increases for extreme head positions back and front compared to the neutral position from 1.5% to 10% and 3%, respectively. For left and right positions of the head movement, the error of prediction was approximately the same as the neutral position: from 1.5% in neutral to 1.2%, and 1.1% for right and left directions, respectively.

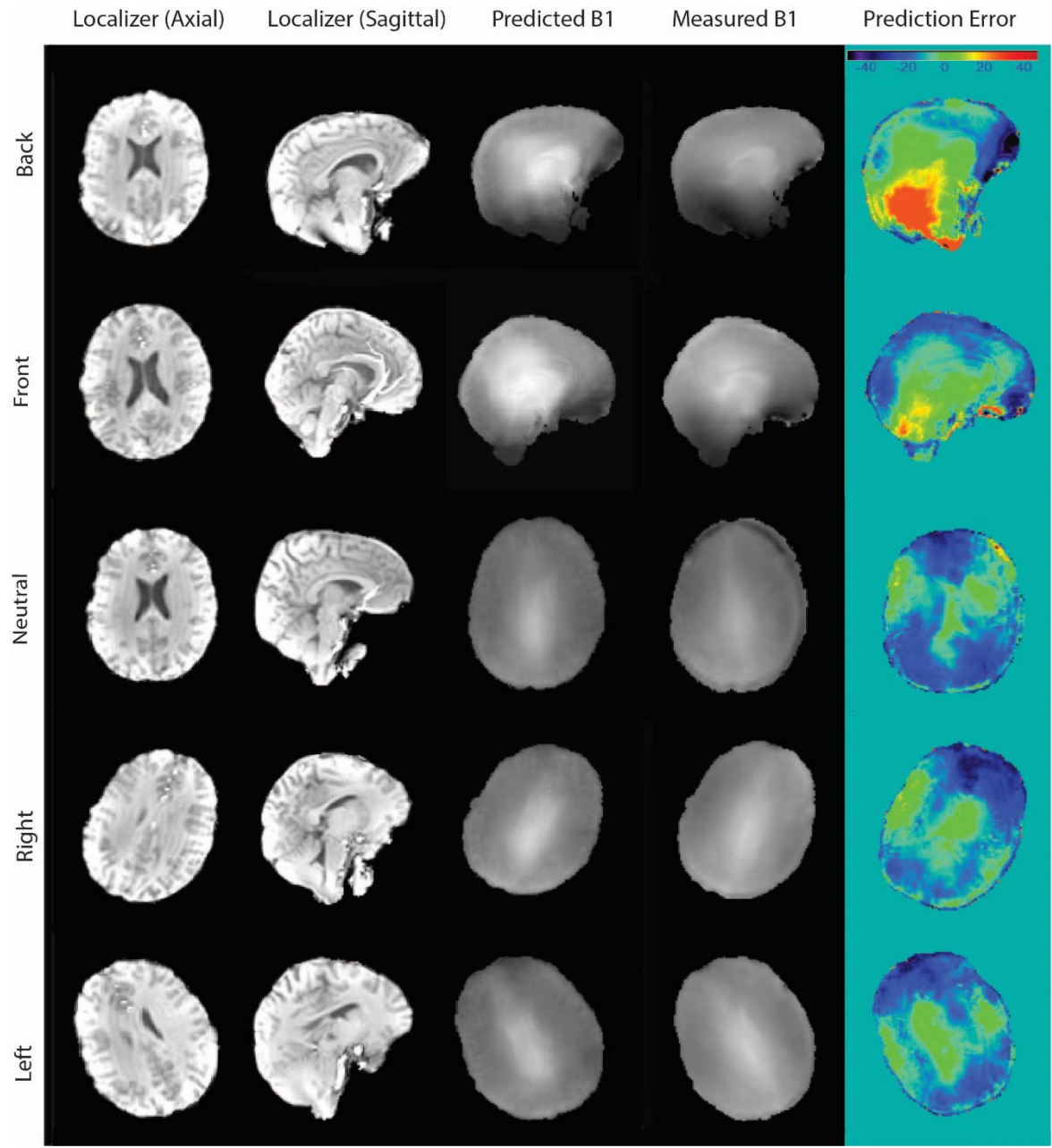

*Figure 4.* The prediction performance of the trained CNN was tested for five different scenarios of head positioning, as shown for one participant (22-year-old male). The words Back, Front, Neutral, Right, and Left refer to the direction of head rotation, as can be seen in the localizer images.

The mean and standard deviation of the scale factors are shown in Figure 5, 6, and 7 calculated in 10 participants for the transverse, coronal, and sagittal orientations, respectively.

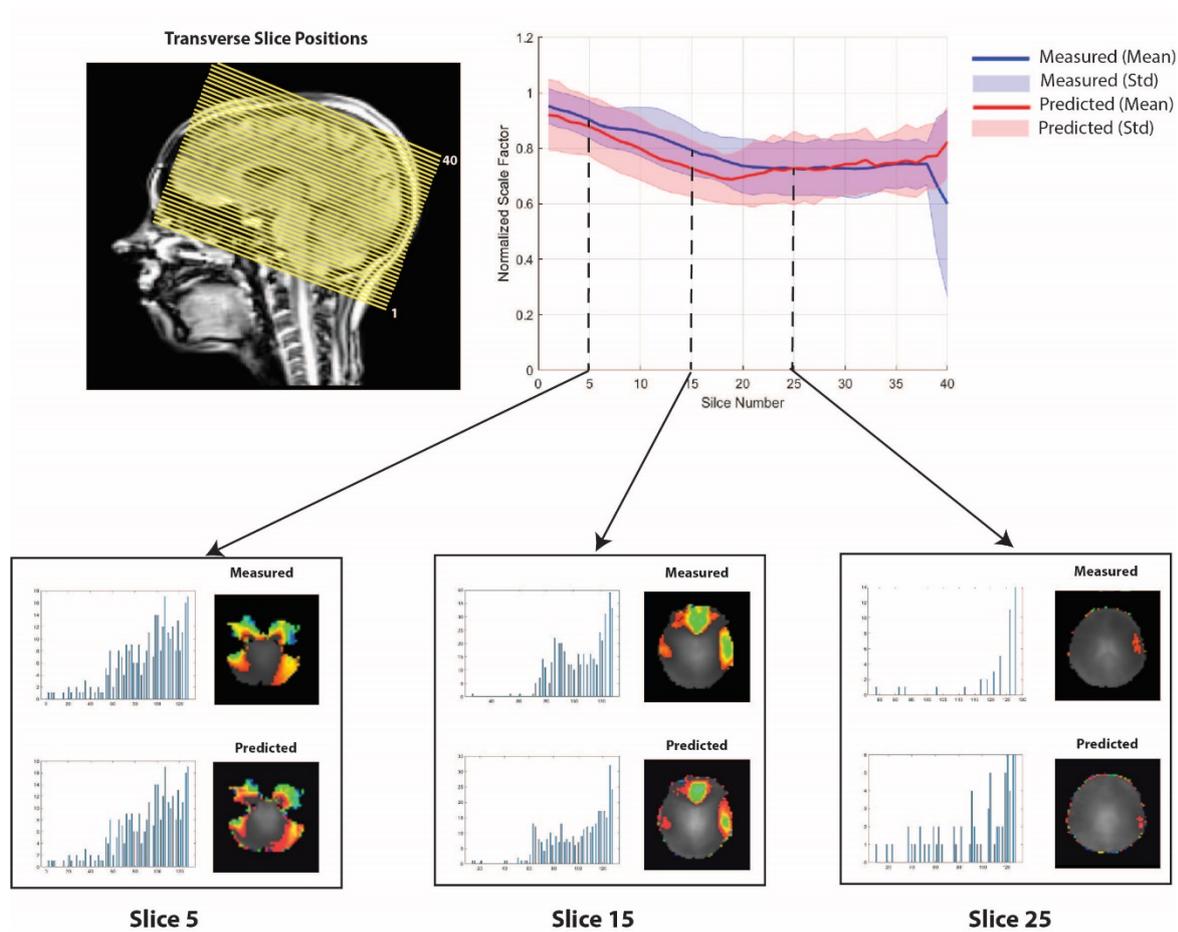

*Figure 5.* Shows the slice positions for transverse imaging and the computed scale factors based on both measured (blue) and predicted (red) $B_1^+$ maps. The solid lines are the mean values and the shaded area represents the standard deviation among 10 participants. The histogram and the map of the $B_1^+$ values that are lower than the adiabatic threshold are shown for three exemplary slices (5, 15, and 25) for both measured and predicted B1 maps

For the transverse orientation (Figure 5), the sequence was using the full power for the adiabatic inversion pulse for the slices in the base of the skull and less power was used in the superior slices. The spatial maps and the histogram of the $B_1^+$ values are shown for the regions where the adiabaticity was not achieved for three exemplary slices (Figure 5). The base of the skull and the temporal lobe of the brain (e.g. slice 5 and 15) contain areas where the adiabatic condition did not hold for the inversion pulse. Thus, the inversion pulse power is scaled to one for these slices. In contrast, around slice 25, scale factors are reduced significantly, as these slices are not containing critical regions with low $B_1^+$ amplitudes. The estimated $B_1^+$ maps predicted the adiabatic pulse power with respect to the adiabaticity of the pulse and the scaling pattern is similar to the one calculated by the measured B1 maps.

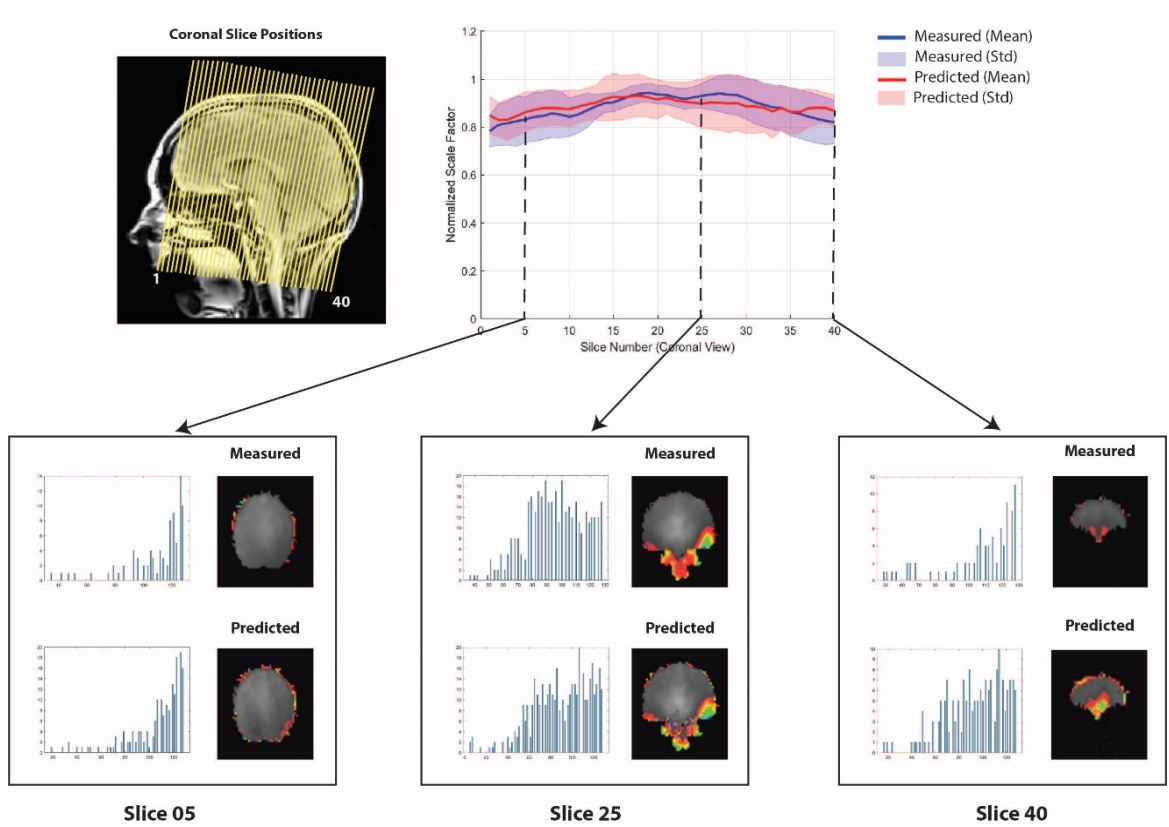

*Figure 6*. Shows the slice positions and the scale factors for coronal orientation calculated based on both measured (blue) and predicted (red) $B_1^+$ maps. The solid lines are the mean values and the shaded area represents the standard deviation among ten participants. The histogram and the spatial map of the $B_1^+$ values that are lower than the value required for adiabaticity of the pulse are shown for three exemplary slices (5, 25, and 40.)

For the coronal orientation (Figure 6), there are only slight discrepancies between the scale factors based on the predicted B1 maps and those based on measured $B_1^+$ maps and they follow the same trend. Acquiring the slices in the coronal orientation, the anterior (e.g. slice 5) and the posterior slices (e.g. slice 40) contain parts of the frontal and occipital lobe, respectively. As the $B_1^+$ inhomogeneity is of less concern in those areas, the power of the pulse is scaled down. At the center of the field-of-view, slices include the base of the skull inferiorly (e.g. slice 25 in Figure 6) and there is a significant area affected by $B_1^+$ inhomogeneity and the power of the pulse is increased.

For the sagittal orientation, shown in **Figure 7**, almost all slices are covering either the temporal lobe or the base of the skull. Nonetheless, some down-scaling of the pulse power is discernible for the slices at the beginning and the end of the field-of-view. The spatial map and the histogram for the $B_1^+$ values that are lower than the value required for adiabaticity of the inversion pulse are shown for slice 4, 20, and 36. Generally, the pulse power is scaled up at the mid-slices (e.g. slice 20) that cover the base of the skull and scaled down at the other slices (e.g. 4 and 36). Although slices 4 and 36 include the temporal lobe of the brain, the severity of the adiabaticity violation for these slices is less than the mid-slices and the pulse power is scaled down slightly.

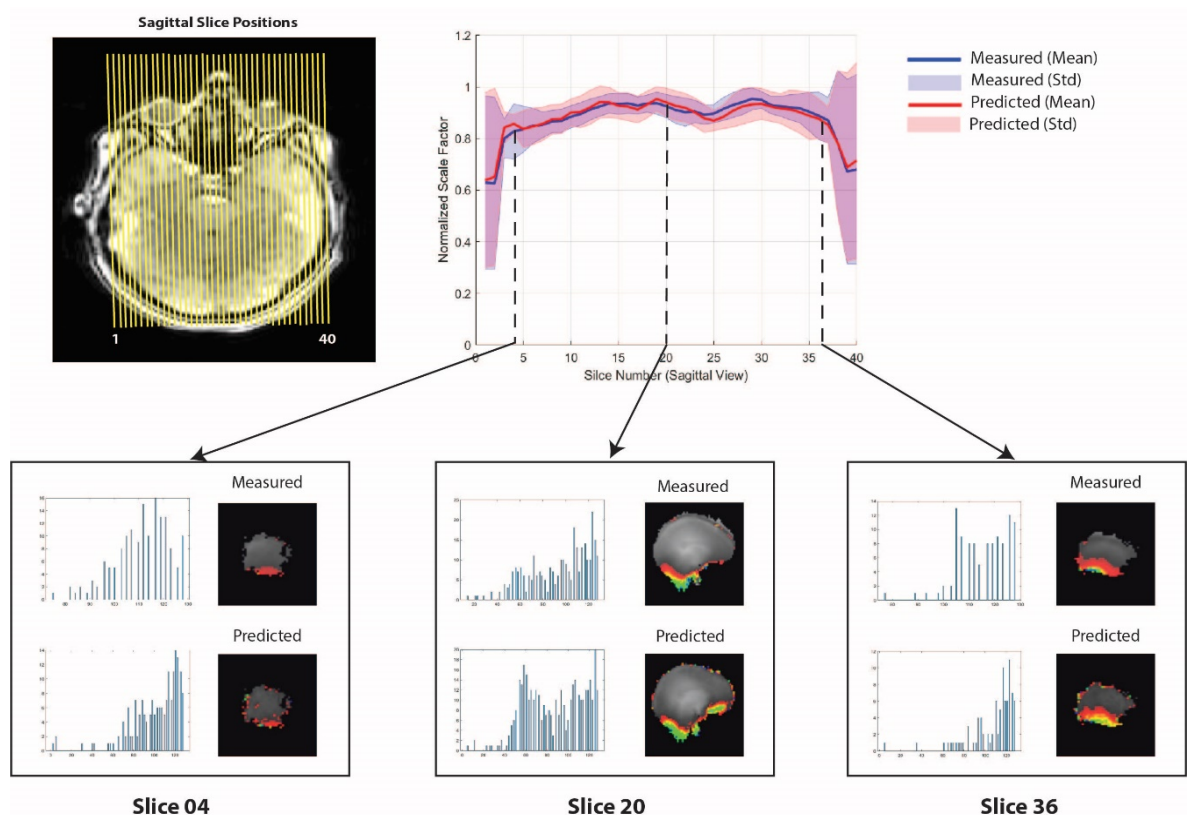

**Figure 7.** Slice positions and scale factors for sagittal orientation calculated based on measured (blue) and predicted (red) $B_1^+$ maps. The solid lines are the mean values and the shaded area represents the standard deviation in ten participants. The histogram and the spatial map of the $B_1^+$ values that are lower than the adiabatic threshold are shown for three exemplary slices (4, 20, and 36).

The SAR reduction experiment was done for six subjects (3M/3F) in transverse, sagittal and coronal slice orientations to investigate the amount of SAR reduction given the predicted slice-by-slice scale factors for TR-FOCI pulse power.

The SAR reduction index for the transverse orientation was calculated using equation 6 for six participants as 43% $\pm$ 6.3% and 42.6% $\pm$ 4.8% corresponding to measured and predicted scale factors, respectively. The standard non-scaled FLAIR acquisition had 140% SAR in the scanners' look ahead monitor. On the other hand, our slice-by-slice scaled mode of FLAIR acquisition resulted in 97% $\pm$ 5.4% and 98% $\pm$ 3.2 % SAR percentages for scale factors computed based on measured and predicted scale factors, respectively. The results show that the proposed SAR reduction strategy reduced the total scan time by 27% in transverse orientation.

The SAR reduction for coronal orientation was computed using equation 6 as 17% ± 4.6% and 15.9% ± 3% for measured and predicted scale factors, respectively. The coronal standard non-scaled FLAIR images were acquired at the SAR level of 140%. Acquisition in our slice-by-slice scaled mode reduced the SAR to 119% ± 3.47 % and 119% ± 3.1 % using the scale factors calculated from measured and predicted $B_1^+$ maps, respectively. The total scan time was reduced by 10.47% and 11.3% for measured and predicted scale factors, respectively.

For the sagittal orientation, the SAR reduction index was calculated as 23.17% ± 6.4 % and 19.36% ± 3.2 % for measured and predicted scale factors, respectively. Non-scaled FLAIR acquisition for sagittal orientation was performed at the SAR level of 140%. In slice-by-slice scaled mode, the FLAIR images were acquired at SAR level of 116% ± 6.5 % and 120% ± 3.5% for scale factors calculated using measured and predicted B1 maps, respectively.

**Table 1**. Results of SAR and scan time reduction for FLAIR acquisitions in two different modes (non-scaled and scaled).

| Orientation | Scaling Basis | SAR reduction index (%) | SAR Value (%)* | Measured SAR reduction (%) | Delay Time (sec) | Scan Time Reduction (%) |
|---|---|---|---|---|---|---|
| Transverse | Non-Scaled | -- | 140% | -- | 91.3 ± 2.21 | -- |
| | Measured | 43.8% ± 6.3% | 97% ± 5.4 % | 43.4% ± 6.15 % | 1 ± 1.52 | 27.45% |
| | Predicted | 42.63% ± 4.8% | 98% ± 3.2 % | 38.2% ± 4.16 % | 1.33 ± 1.97 | 27.45% |
| Coronal | Non-Scaled | -- | 140% | -- | 90.16 ± 1.95 | -- |
| | Measured | 17.01% ± 4.6% | 119% ± 3.47 % | 20.85% ± 3.47 % | 52.66 ± 9.55 | 10.47% |
| | Predicted | 15.88% ± 3% | 119% ± 3.1 % | 20% ± 3.1 % | 55 ± 8.52 | 11.3% |
| Sagittal | Non-Scaled | -- | 140% | -- | -- ᵗ | -- |
| | Measured | 23.17% ± 6.4% | 116% ± 6.5 % | 23.59% ± 6.5 % | 45.16 ± 18 | 16.13% |
| | Predicted | 19.36% ± 3.2% | 120% ± 3.5 % | 19.18% ± 3.51 % | 57.33 ± 9.65 | 13.1% |

* The value read from SAR look ahead monitor

ᵗ The SAR exceed was compensated by reducing the field of view

Acquisitions at the SAR level of higher than 100% were feasible by the scanner by adding a delay time after the scan to avoid cumulative energy deposition of the pulse exceeding the limits. **Table 1** summarizes all six experiments of FLAIR acquisitions, each in nine different acquisitions (3 different slice orientations with non-scaled, scaled from measured $B_1^+$ maps and scaled from predicted $B_1^+$ maps). The delay times enforced by the scanner software after each acquisition are also reported in Table 1. The results show that the scaled mode acquisition of the FLAIR sequence resulted in SAR reductions, which led to the total scan time reduction of 27%, 11%, and 15% for transverse, coronal, and sagittal orientations, respectively.

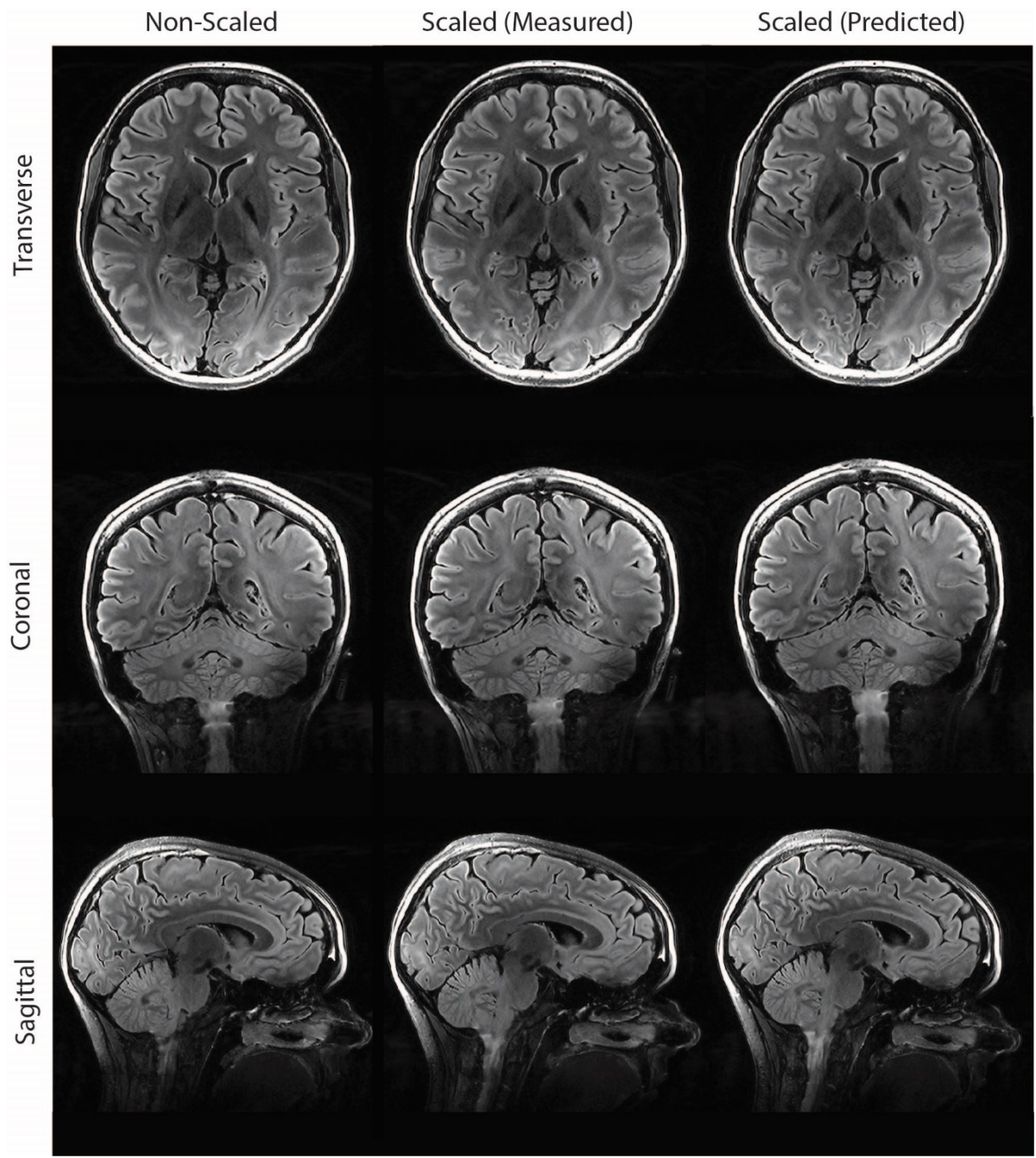

*Figure 8. Shown are the FLAIR images in nine different acquisitions. The first column is in non-scaled mode, from top to bottom, in transverse, coronal, and sagittal orientations. The second and third column shows the FLAIR images acquired in slice-by-slice mode using the scale factors calculated from measured and predicted $B_1^+$ maps, respectively.*

**Figure 8** shows the image quality of FLAIR images acquired in different modes, non-scaled and scaled. The FLAIR images in nine different modes show that the scaled mode acquisition does not affect the CSF suppression or image quality.

## 4. Discussion

In this study, we assumed that the power of the adiabatic inversion pulse used in the FLAIR sequence could be adapted based on the $B_1^+$ profile of each slice to reduce SAR requirement without affecting image quality. We used a deep convolutional neural network to estimate the $B_1^+$ profile, simulated Bloch equations to calculate the required absolute $B_1^+$ value for adiabaticity, and accordingly calculated the adiabatic inversion pulse power in a slice-by-slice fashion. Thus, $B_1^+$ inhomogeneity mitigation and SAR reduction were achieved at the same time to make T$_2$-FLAIR imaging up to 27 percent faster with full brain coverage at 7T.

The challenges at ultra-high-field have been previously studied and our proposed solution adds to these concepts: Zwanenburg *et al.* used adiabatic inversion pulses for FLAIR at 7 T to improve transmit B1 homogeneity and showed acceptable image quality (7), but no additional strategy for SAR reduction was proposed yet. Visser *et al*. used a magnetization preparation module consisting of four adiabatic refocussing pulses for 3D FLAIR at 7T, improving CNR but exacerbating the SAR issue of the sequence (29). O'Brien et al. used high permittivity dielectric pads for the MP2RAGE sequence as a complementary technique to adiabatic inversion allowing to lower the power of the pulse and reduce the SAR (18). Parallel transmit (PTx)-based solutions were recently proposed for implementing FLAIR at 7T using dynamic RF shimming or $k_T$-point pulses (19, 20, 30, 31). However, these proposed solutions need additional hardware for implementation; the dielectric pads are not desirable for efficient clinical workflows and multi-channel transmit coils are not yet widely available. In contrast, our proposed technique was capable of improving $B_1^+$ homogeneity and reducing the SAR at the same time using more readily available single channel transmit coil, without additional hardware or scans that may disturb the clinical workflow.

As the inversion of the spins was the key feature of the sequence and was hindered by the B1 transmit inhomogeneity of the inversion pulse, the adiabatic TR-FOCI pulse was used to assure spin inversion. In the standard non-scaled mode, where the same pulse power was used for every slice, the power was increased to the threshold and the adiabaticity was achieved for the regions severely affected by $B_1^+$ inhomogeneity such as the base of the skull and the temporal lobes. Those slices that were not severely affected by $B_1^+$ inhomogeneity were over-driven by the pulse power, which unnecessarily increased the SAR load of the sequence. Our proposed deep learning-based technique showed we can predict the adiabatic inversion pulse power using the inherent $B_1^+$ inhomogeneity observed in the AutoAlign 3D to sufficiently adapt the inversion pulse on a slice-by-slice basis without the need for additional scan time.

The versatility of the technique with respect to the slice orientation was illustrated for the three primary imaging orientations (transverse, sagittal, and coronal). In all three orientations, the adiabatic pulse power was reduced for the slices including the base of the skull or the temporal lobes of the brain. Bloch equation simulations for the RF pulse achieved robustness of the scale factor calculation. By knowing the adequate power required for adiabaticity, it was ensured that the downscaling of the pulse power never violated the adiabaticity of the TR-FOCI pulse for each slice. Therefore, by reducing SAR, the CSF suppression remained intact and the image quality was not affected.

Results in Table 1 illustrate that our proposed technique reduced SAR in all three primary orientations. Using the technique for FLAIR imaging in transverse orientation reduced the SAR by as much as 40% and completely removed the delay time, while imaging in sagittal and coronal orientations reduces SAR by 21% and 20% still enforcing a delay time of 51 and 53 seconds due to the SAR excess, respectively.

Despite slight discrepancies in the scale factors calculated based on measured $B_1^+$ maps and predicted $B_1^+$ maps, the scaling patterns followed the same trend. This similar trend showed that the adiabatic pulse power prediction strategy was robust to the under- or over-estimations of the $B_1^+$ values caused by prediction errors of the convolutional neural network. The standard deviation of the scale factors among subjects imply that the scale factors are subject-dependent and highlights the need for the DL-based approach to predict the B1 map for each subject prior to imaging and that a simple heuristic model would not provide the robustness needed for a seamless clinical workflow.

While we have developed this technique with a specific sequence and UHF MRI in mind, the principle is not restricted to T$_2$-FLAIR at 7T. MRI sequences with adiabatic pulses that are acquired in a 2D multi-slice mode that are SAR restricted even at lower field would potentially profit from this approach. We also think that the proposed technique is promising for streamlining the workflow of parallel transmission (pTx) techniques, as it might be possible to extract sufficiently accurate transmit B1 information needed for signal optimization from the localizer scan (32).

In conclusion, we showed that it is possible to reduce SAR requirements of the T$_2$-FLAIR sequence at 7T by tailoring the pulse power of an adiabatic TR-FOCI pulse for each slice based on predicting $B_1^+$ from a 3D localizer scan using a convolutional neural network. This enabled a reduction in scan time for the T$_2$-FLAIR sequence without any additional hardware by using readily available single channel transmit coils.

## Declarations of interest

Authors KO, VV, AR, YT, MB and SB are co-inventors of a patent "Prediction of calibration maps from standard localizers using deep convolutional neural networks" filed on 20th Apr 2017 (62/487,629).

## Acknowledgements

We thank the participants involved in this study. MB acknowledges funding from Australian Research Council Future Fellowship grant FT140100865, and VV from NHMRC Project Grant AP1104933. SB acknowledges funding from UQ Postdoctoral Research Fellowship grant and an NVIDIA Hardware Seed Grant. The authors acknowledge the facilities and scientific and technical assistance of the National Imaging Facility, a National Collaborative Research Infrastructure Strategy (NCRIS) capability, at the Centre for Advanced Imaging, the University of Queensland.